\documentclass[11pt,a4paper]{article}
\usepackage{epsfig}
\usepackage{psfrag}
\usepackage{enumitem}
\usepackage{latexsym}
\usepackage{booktabs}
\usepackage{indentfirst}
\usepackage{fancyhdr}
\usepackage{amsfonts, amssymb}
\usepackage[centertags]{amsmath}
\usepackage{pifont}
\usepackage{cite}
\usepackage{bbold}
\usepackage{color}
\usepackage[footnotesize]{caption2}
\usepackage{graphics}
\usepackage[center,footnotesize,hang]{subfigure}
\usepackage{url}
\usepackage[comma,square,numbers,sort]{natbib}
\usepackage{array}
\usepackage{subfigure}

\newcommand{\Dmq}{\Delta m^2}
\newcommand{\Nuc}[2][]{{\ensuremath{\ifthenelse{\equal{#1}{}}{}{\mbox{}^{#1}}\text{#2}}}}

\newcommand{\summ}{\Sigma_i\, m_{\nu_i}}

\def\be{\begin{equation}}
\def\ee{\end{equation}}
\def\bc{\begin{center}}
\def\ec{\end{center}}
\def\bea{\begin{eqnarray}}
\def\eea{\end{eqnarray}}

\newcommand{\ba}{\begin{array}{c}}
\newcommand{\bad}{\begin{array}{ccc}}
\newcommand{\ea}{\end{array}}

\def\nn{\nonumber}

\DeclareMathOperator{\diag}{Diag}
\newcommand{\meff}{\mbox{$\left|  \langle \!\,  m  \,\!  \rangle \right|$ }}
\newcommand{\betabeta}{\mbox{$(\beta \beta)_{0 \nu}  $}}

\begin{document}

\begin{titlepage}
\hfill{RM3-TH/14-1}
  \vskip 2.5cm
   \begin{center}
    {\Large\bf Two-zero Majorana textures in the light of the Planck results}
   \end{center}
  \vskip 0.2  cm
   \vskip 0.5  cm
  \begin{center}
\vskip .2cm
{\large Davide Meloni}~\footnote{e-mail address: meloni@fis.uniroma3.it}
\\
\vskip .1cm
Dipartimento di Matematica e Fisica, Universit\`a di Roma Tre
\\
INFN, Sezione di Roma Tre, \\
Via della Vasca Navale 84, I-00146 Rome, Italy\\
\vskip .2cm
{\large Aurora Meroni}~\footnote{e-mail address: ameroni@fis.uniroma3.it}
\\
\vskip .1cm
Dipartimento di Matematica e Fisica, Universit\`a di Roma Tre
\\
Via della Vasca Navale 84, I-00146 Rome, Italy\\
INFN, Laboratori Nazionali di Frascati, \\
Via Enrico Fermi 40, I-00044 Frascati, Italy
\\
\vskip .2cm
{\large Eduardo Peinado}~\footnote{e-mail address: epeinado@lnf.infn.it}
\\
\vskip .1cm
INFN, Laboratori Nazionali di Frascati, \\
Via Enrico Fermi 40, I-00044 Frascati, Italy
\end{center}
\vskip 0.7cm

\begin{abstract}
The recent results of the Planck experiment put a stringent
constraint on the sum of the light neutrino masses, $\Sigma_i m_i <
0.23$ eV (95\% CL). On the other hand, two-zero Majorana mass matrix
textures predict strong correlations among the atmospheric angle
$\sin^2 \theta_{23}$ and $\Sigma$. We use the Planck result to show
that, for the normal hierarchy case, the texture with vanishing
$(2,2)$ and $(3,3)$ elements is ruled out at a high confidence
level; in addition, we emphasize that a future measurement of the
octant of $\theta_{23}$ (or the 1$\sigma$ determination of it based
on recent fit to neutrino data) will put severe constraint on the
possible structure of the Majorana mass matrix. The implication of
the above mentioned correlations for neutrinoless double
$\beta$-decay  are also discussed, for both normal and inverted
orderings.
\end{abstract}
\end{titlepage}

\section{Introduction}
The recent results from solar, atmospheric and reactor neutrino
experiments have proven that neutrinos are massive, with at least
two of them being non-relativistic today. Beside the values of the
two squared mass differences and mixing angles
\cite{Tortola:2012te,Fogli:2012ua,GonzalezGarcia:2012sz}\footnote{In
this paper we will use the fit results summarized in Table
\ref{tab:neu} of \cite{GonzalezGarcia:2012sz}; the outcome of our
analysis does not change if other data set are used.}, recently the
Planck Collaboration \cite{Ade:2013lta} has released a quite
stringent upper bound on the sum of the active neutrino masses:
 \be \Sigma \equiv \summ < 0.23 \,\mbox{eV}\qquad  \mbox{95\% CL}
\label{eq:PlanckBAO}\,.\ee

This limit has been obtained assuming three species of degenerate
massive neutrinos and a $\Lambda$CDM model and it stems from the
combination of the Planck temperature power spectrum with a WMAP
polarization low-multipole likelihood at $\ell\leq$23 and the Baryon
Acoustic Oscillation data. The above upper limit can be converted
into a limit on the absolute scale of neutrino masses that reads
$m_{min}\lesssim 0.07 $ eV.

In addition, the South Pole Telescope Collaboration released a fit
analysis that indicates a preferred value for the sum of the light
neutrino, $\Sigma=0.32\pm0.11$ eV, with a 3$\sigma$ detection of
positive neutrino mass in the range $[0.01, 0.63]$ eV at 99.7 \%
C.L. \cite{Hou:2012xq}. This result will be 
tested by the KATRIN experiment which is expected to have a
sensitivity around $0.2$ eV \cite{katrinexp}. All this has revived
some interest in connection with theories of neutrino mass
generation predicting a quasi-degenerate (QD) spectrum for the light
active neutrinos.

The bounds and values described above however are not definitive and
other  forthcoming observations will test them. The EUCLID survey
\cite{Laureijs:2011mu}  will be able most likely  to measure the
neutrino mass sum at the 0.01 eV level of precision combining the
data with measurements of the CAMB anisotropies from the Planck
mission. Such an outstanding precision will be able to unveil the
absolute scale of neutrino masses \cite{Hamann:2012fe}, being the
minimum of the sum compatible with current neutrino oscillation data
$\Sigma_{min}=5.87 \times10^{-2}$ eV for a normal hierarchical (NO)
spectrum and $\Sigma_{min}=9.78 \times10^{-2}$ eV for an inverted
hierarchical spectrum (IO).

\begin{table}[h!]\centering
  \footnotesize
  \begin{tabular}{@{}lcc@{}}
    \toprule
    & best fit $\pm 1\sigma$ & $3\sigma$ range
    \\
\midrule
    \rule{0pt}{4mm}
    $\sin^2\theta_{12}$
    & $0.306_{-0.012}^{+0.012}$ & $0.271 \to 0.346$
    \\[1.5mm]
    $\sin^2\theta_{23}$
    & $0.446_{-0.007}^{+0.007} \oplus 0.587_{-0.037}^{+0.032}$ & $0.366 \to 0.663$
    \\[1.5mm]
    $\sin^2\theta_{13}$
    & $0.0229_{-0.0019}^{+0.0020}$ & $0.0170 \to 0.0288$
    \\[1.5mm]
    $\Dmq_{12} \; [10^{-5}\text{ eV}^2]$
    & $7.45_{-0.16}^{+0.19}$ & $6.98 \to 8.05$
    \\[3mm]
     $\Dmq_{31} \; [10^{-3}\text{ eV}^2] $ (NO)
    & $+2.417_{-0.013}^{+0.013}$ & $+2.247 \to +2.623$
     \\[3mm]
     $\Dmq_{32} \; [10^{-3}\text{ eV}^2]$ (IO)
    & $-2.410_{-0.062}^{+0.062}$ & $-2.602 \to -2.226$
    \\[3mm]
    \bottomrule
  \end{tabular}
  \caption{\it Three-flavour oscillation parameters from the fit to global
    data after the TAUP 2013 conference from \cite{GonzalezGarcia:2012sz}.
The reactor fluxes have been left free in the fit and short
    baseline reactor data (RSBL) with $L \lesssim 100$~m are included.}
  \label{tab:neu}
\end{table}

This arises the question on whether the new Planck result could be used to
infer some general properties of the neutrino mass matrix and, in
particular, to discard some of the neutrino mass textures compatible
with the oscillation data. To be predictive, one generally expect
that some of the elements of the neutrino mass matrix must be
vanishing or strongly related. Here we focus on the first class of
matrices; among the Majorana mass matrices with two-zero entries, we
restrict our analysis to the five models still compatible with the
data and having a non-vanishing Majorana effective mass \meff  for
the neutrinoless double $\beta$-decay (\betabeta-decay); according
to \cite{preafter,Fritzsch:2011qv}, from which we also adopt here the
nomenclature, the interesting textures are:

\be \label{eq:textu}
\begin{split}
B^{}_1: ~~ \left(
\begin{matrix}
a_1 & a_2 & 0 \cr a_2 & 0 &
a_3 \cr 0 & a_3 & a_4 \cr
\end{matrix}\right) \, , & \qquad
B^{}_2: ~~
\left(
\begin{matrix}
a_1 & 0 & a_2 \cr 0 & a_3 & a_4 \cr
a_2 & a_4 & 0 \cr
\end{matrix}\right) \, , ~~~~~
\qquad
B^{}_3: ~~ \left(\begin{matrix}
a_1 & 0 & a_2 \cr 0 & 0 &
a_3 \cr a_2 & a_3 & a_4 \cr
\end{matrix} \right) \, ,
\\ \\
& B^{}_4: ~~ \left(\begin{matrix}a_1 & a_2 & 0 \cr a_2 & a_4 & a_3
\cr 0 & a_3 & 0 \cr
\end{matrix}\right) \, , ~~~~~
C: ~~ \left(
\begin{matrix} a_1 & a_2 & a_3 \cr a_2 & 0 &
a_4 \cr a_3 & a_4 & 0 \cr\end{matrix} \right) \; .
\end{split}
\ee

All these textures give rise to a QD neutrino mass spectra and can
therefore be constrained or even ruled out by current and
forthcoming experimental results coming from cosmology and from the
next generation of \betabeta-decay experiments.

In the present work we want to study the compatibility of these
patterns with the present data using the best fit values of Table
\ref{tab:neu} and show which ones are already saturating the current
bounds obtained by cosmological observations and other source of
data such as \betabeta-decay limits, taking into account the octant
degeneracy of $\theta_{23}$. It is worth noticing that a recent
global fit analysis has been performed \cite{Capozzi:2013csa} in
which possible hints related to the value of the Dirac phase
$\delta$ responsible for CP violation in the lepton sector and to
the octant of $\theta_{23}$ in NO emerge. This up-to-date analysis
comes from the combination of different kind of data, e.g. reactor
and accelerator experiments, and of their interplay with solar and
atmospheric data and seems to indicate a preferred value for the
Dirac phase $\delta\sim 3\pi/2$ ---specifically $\delta=1.39 (1.35)
\pi$ for NO (IO) and for the octant of $\theta_{23}$ in the first
octant for NO (for IO two independent regions still exist at
1$\sigma$). In what follows we will comment more on these
indications; however, our numerical analysis will be mainly based on
the results of the global fit performed in
\cite{GonzalezGarcia:2012sz} where no preferred octant for
$\theta_{23}$ in NO emerged from their global analysis, since we
want to study the dependence of the textures in eq. (\ref{eq:textu})
on the two possibilities for $\theta_{23}$. We will use the Planck
result to show that, for  normal ordering, the texture $C$ is ruled
out at a high confidence level and we will show that a future
measurement of the octant of $\theta_{23}$ will put sever constraint
on the possible structure of the Majorana mass matrix. We will show
also the implication of the above mentioned correlations for
\betabeta-decay, for both normal and inverted orderings.

\section{General Approach and Results}
The mixing of three light massive neutrinos $\nu_i$, $i=1,2,3$, in
the weak charged lepton current  is described by the Pontecorvo,
Maki, Nakagawa, Sakata (PMNS) $3\times 3$ unitary mixing matrix,
$U_{\rm PMNS}$. In the  standard parametrisation \cite{PDG10},
$U_{\rm PMNS}$ is expressed in terms of the solar, atmospheric and
reactor neutrino mixing angles $\theta_{12}$, $\theta_{23}$ and
$\theta_{13}$, respectively, and one Dirac - $\delta$, and two
Majorana \cite{oscillphases} - $\alpha_{21}$ and $\alpha_{31}$, CP
violating (CPV) phases:
\be
 U_{\text{PMNS}} \equiv U = V(\theta_{12},\theta_{23},\theta_{13},\delta)\,
Q(\alpha_{21},\alpha_{31})\,, \label{eq:UPMNS} \ee
%
where
\be V = \left(
     \begin{array}{ccc}
       1 & 0 & 0 \\
       0 & c_{23} & s_{23} \\
       0 & -s_{23} & c_{23} \\
     \end{array}
   \right)\left(
            \begin{array}{ccc}
              c_{13} & 0 & s_{13}e^{-i \delta} \\
              0 & 1 & 0 \\
              -s_{13}e^{i \delta} & 0 & c_{13} \\
            \end{array}
          \right)\left(
                   \begin{array}{ccc}
                     c_{12} & s_{12} & 0 \\
                     -s_{12} & c_{12} & 0 \\
                     0 & 0 & 1 \\
                   \end{array}
                 \right)\,,
\label{eq:V} \ee
%
and we have used the standard notation $c_{ij} \equiv
\cos\theta_{ij}$, $s_{ij} \equiv \sin\theta_{ij}$ and
\be Q =  \diag(1, e^{i \alpha_{21}/2}, e^{i \alpha_{31}/2})\,.
\label{Q} \ee
Therefore the Majorana mass matrix of neutrinos in the flavour basis
can be written as:

\be M_\nu=M_\nu^T= U\, diag(m_1,m_2,m_3)\,U^T \ee where $m_1$,
$m_2$, $m_3$ can be chosen real and positive. The data coming from
oscillation experiments constrain  the neutrino mass spectrum to be
either i) normal ordered (NO): $m_1 < m_2 < m_3$, $\Delta m^2_{31}
>0$, $\Delta m^2_{21} > 0$, $m_{2(3)} = (m_1^2 + \Delta
m^2_{21(31)})^{1/2}$; or ii)  inverted ordered (IO): $m_3 < m_1 <
m_2$, $\Delta m^2_{32}< 0$, $\Delta m^2_{21} > 0$, $m_{2} = (m_3^2 +
\Delta m^2_{23})^{1/2}$, $m_{1} = (m_3^2 + \Delta m^2_{23} - \Delta
m^2_{21})^{1/2}$. For a range of the lightest neutrino mass $\gtrsim
0.10$ eV  the neutrino mass spectrum is said to be quasi-degenerate with $m_1 \cong m_2 \cong m_3$ , $m_j^2 \gg |\Delta
m^2_{31(32)}|$ with $j=1,2,3$ while  it is NO
if  $m_1 \ll m_2 < m_3$, $m_2 \cong (\Delta m^2_{21})^{1/2} $, $m_3
\cong (\Delta m^2_{31})^{1/2} $  or IO if
$m_3 \ll m_1 < m_2$, with $m_{1,2} \cong |\Delta m^2_{32}|^{1/2}$.

The matrix $M_\nu$  is described altogether by nine parameters: the
three masses, three mixing angles and three CPV phases. The textures
in eq. (\ref{eq:textu}) are particularly interesting since the two
zero entries impose four independent constraints on the mixing
parameters. Considering the experimental values on the three mixing
angles $\theta_{ij}$, and the two squared mass differences $\Delta
m^2_{21}$ and $\Delta m^2_{31(32)}$ we have in total nine relations
that allow to fix all the parameters of $M_\nu$, including the CPV
phases $\delta$, $\alpha_{21}$ and $\alpha_{31}$. Details on this
procedure can be found, among others, in refs.
\cite{preafter,Fritzsch:2011qv}.
We have updated the predictions
from the five textures on the values of the neutrino masses, CPV
phases, the Majorana effective mass \meff in the  \betabeta-decay
and the sum of the neutrino masses $\Sigma$, using the best fit
values quoted in Table \ref{tab:neu}. Our results are given in
Tables \ref{tab:numericsNO} and \ref{tab:numericsIO}, where we list
the numerical solutions for the above-mentioned neutrino observables
for NO and IO spectrum, respectively.
\begin{table}
\centering
\renewcommand{\arraystretch}{1.1}
\begin{tabular}{cccccc}
\toprule Texture    & ($m_1 , m_2 , m_3$) \,[eV] & $\pm(\delta,\alpha_{21},\alpha_{31})$ & \meff  \,[eV] & $\Sigma$ \,[eV] \\
\midrule
$B_1 \, (\theta_{23}^{\ell})$ & $(6.27, 6.33, 7.97)\times10^{-2}$    &  $ (91.7, \, 5.61, \, 183.8)^\circ$    & $6.32 \times 10^{-2}$ & 0.206\\
$B_2 \, (\theta_{23}^{u})$   &  $(4.53, 4.61, 6.69)\times10^{-2}$    &  $ (86.96, \, 349.45, \, 172.77)^\circ$    & $4.59 \times 10^{-2}$ & 0.158\\
$B_3\, (\theta_{23}^{\ell})$  & $(6.66, 6.71, 8.28)\times10^{-2}$    &  $ (91.96, \, 355.95, \, 177.15)^\circ$    & $6.71 \times 10^{-2}$ & 0.216\\
$B_4\, (\theta_{23}^{u})$  &  $(4.84, 4.92, 6.90)\times10^{-2}$    &  $ (92.66, \, 6.61, \, 184.65)^\circ$    & $4.90 \times 10^{-2}$ & 0.167\\
 \bottomrule
\end{tabular}
\caption{\it Numerical solutions for several neutrino observables corresponding to the best fits in Table
\ref{tab:neu} for NO. We employ the values of
$\theta_{23}$ given for both octants ($\ell$ lower and $u$ upper
octants). \label{tab:numericsNO}}
\end{table}

\begin{table}
\centering
\renewcommand{\arraystretch}{1.1}
\begin{tabular}{ccccc}
\toprule Texture   & ($m_1 , m_2 , m_3$) \,[eV] & $\pm(\delta,\alpha_{21},\alpha_{31})$ & \meff  \,[eV] & $\Sigma$ \,[eV] \\
\midrule
$B_1\,(\theta_{23}^{u})$ & $(6.68, 6.73, 4.61)\times10^{-2}$    &  $ (90.01, \, 355.10, \, 176.43)^\circ$    & $6.64 \times 10^{-2}$ & 0.180\\
$B_2\,(\theta_{23}^{\ell})$ & $(7.98, 8.03, 6.35)\times10^{-2}$    &  $ (90.08, \, 3.46, \, 182.50)^\circ$    & $7.95 \times 10^{-2}$ & 0.224\\
$B_3\,(\theta_{23}^{u})$  & $(6.89, 6.95, 4.91)\times10^{-2}$    &  $ (90.56, \, 6.44, \, 184.44)^\circ$    & $6.85 \times 10^{-2}$ & 0.188\\
$B_4\,(\theta_{23}^{\ell})$ & $(8.28, 8.32, 6.72)\times10^{-2}$    &  $ (89.71, \, 356.04, \, 177.28)^\circ$    & $8.24 \times 10^{-2}$ & 0.233\\
$C \,(\theta_{23}^{\ell})$  & $(8.35, 8.39, 6.81)\times10^{-2}$    &  $ (72.10, \, 281.40, \, 123.00)^\circ$    & $6.79 \times 10^{-2}$ & 0.235\\
$C \,(\theta_{23}^{u})$  & $(6.39, 6.45, 4.19)\times10^{-2}$    &  $ (118.89, \, 111.28, \, 265.44)^\circ$    & $4.15 \times 10^{-2}$ & 0.170\\
 \bottomrule
\end{tabular}
\caption{\it Numerical solutions  for several neutrino observables corresponding to the best fits in Table
\ref{tab:neu} for IO. We employ the values of
$\theta_{23}$ given for both octants ($\ell$ lower and $u$ upper
octants).\label{tab:numericsIO}}
\end{table}
First of all, from our numerical results it is clear that within the
analyzed textures only QD spectrum is possible  for the three light
active neutrinos. This is particularly important in view of the
recent cosmological results and the next generation of
\betabeta-decay experiments which can strongly constrain or even
rule out the textures under study. For the textures of type B, the
values of the Majorana phases are near CP conserving values
(similarly to what found in \cite{Fritzsch:2011qv}), whereas the
Dirac CP phase $\delta$ for both orderings is approximately
$\pm\pi/2$. Interestingly enough, these values seem to be compatible
with the intriguing indication suggested in \cite{Capozzi:2013csa}.

From Table \ref{tab:numericsNO} and \ref{tab:numericsIO} it is
evident that the NO is possible only for the textures $B_1$ and
$B_3$ ($B_2$ and $B_4$) when the value of $\theta_{23}$ in the lower
(upper) octant is used while IO is possible if  $B_1$ and $B_3$
($B_2$ and $B_4$) are associated with the upper (lower)
$\theta_{23}$ value. Finally, the texture $C$ only allows for IO
using both values of $\theta_{23}$ while the NO in the case of the
current best fit values is not consistent (in other words, as
explained in \cite{Grimus:2004az}, this texture predicts
$\theta_{23}$ to be almost maximal). In addition, in this case the
Dirac CPV phase is drifting apart from its maximal value.

Looking at Tables \ref{tab:numericsNO} and \ref{tab:numericsIO}, we
also have to notice that  some of the B textures, namely
$B_1(\theta_{23}^\ell)$ and $B_3(\theta_{23}^\ell)$ for NO and
$B_2(\theta_{23}^u)$, $B_4(\theta_{23}^u)$ and $C(\theta_{23}^\ell)$
for IO,  correspond to a value of $\Sigma$ that, for the present
best fit values, is near to the current cosmological bound given in
eq. (\ref{eq:PlanckBAO}) and therefore they are now close to be
strongly constrained. Moreover all these textures, independently of
the octant of $\theta_{23}$, have a range for $\meff> 10^{-2}$ eV
(as a consequence of the QD spectrum), which is the range planned to
be exploited by the next generation of $\betabeta$-experiments.

\begin{figure}[h!]
  \begin{center}
 \subfigure
 {\includegraphics[width=7cm]{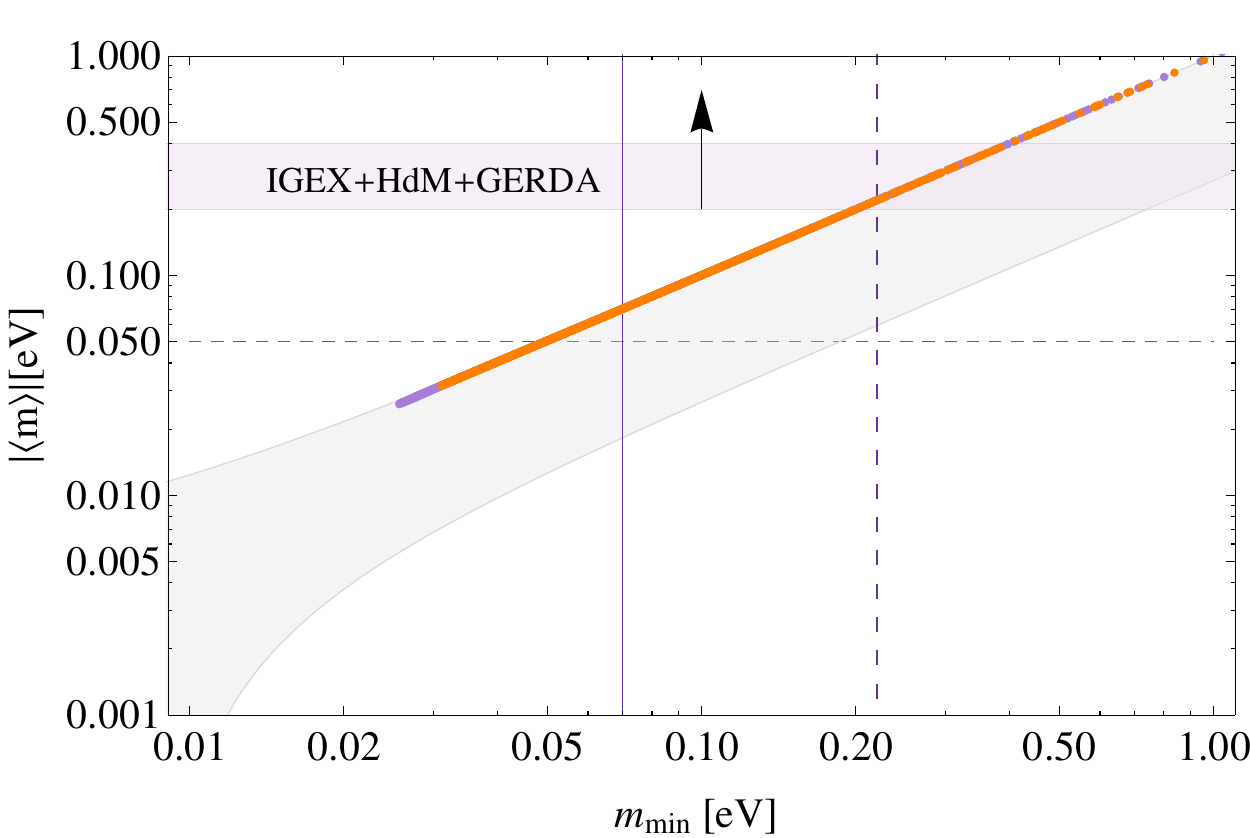}}
 \vspace{5mm}
 \subfigure
   {\includegraphics[width=7cm]{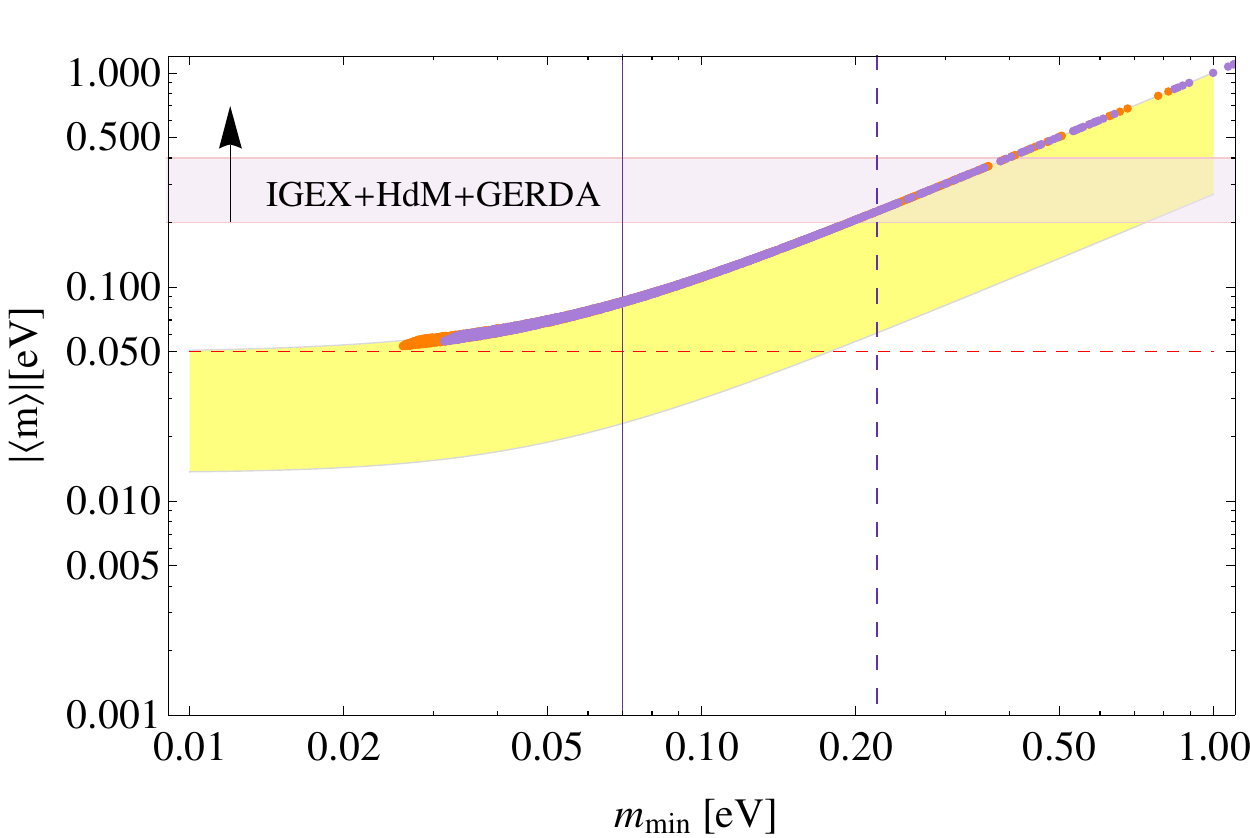}} \\
 {\includegraphics[width=7cm]{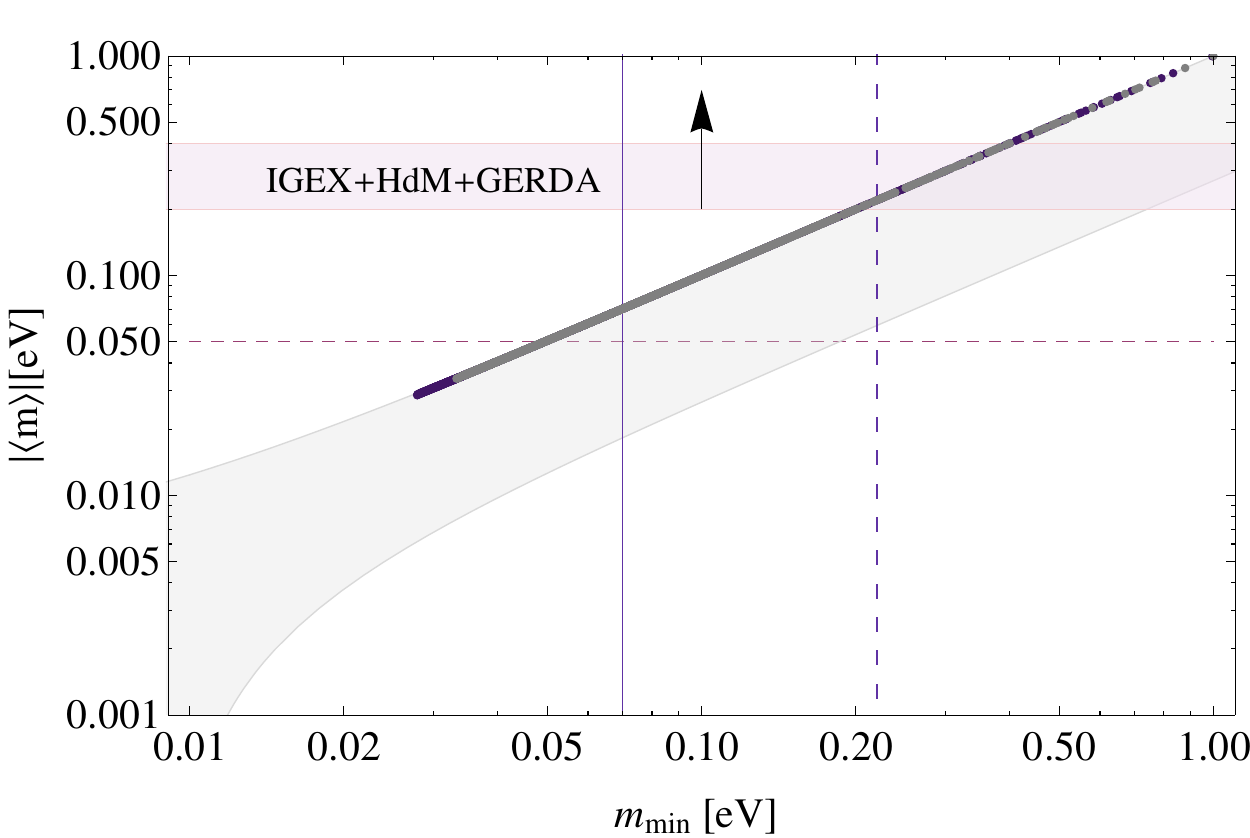}}
 \vspace{5mm}
 \subfigure
   {\includegraphics[width=7cm]{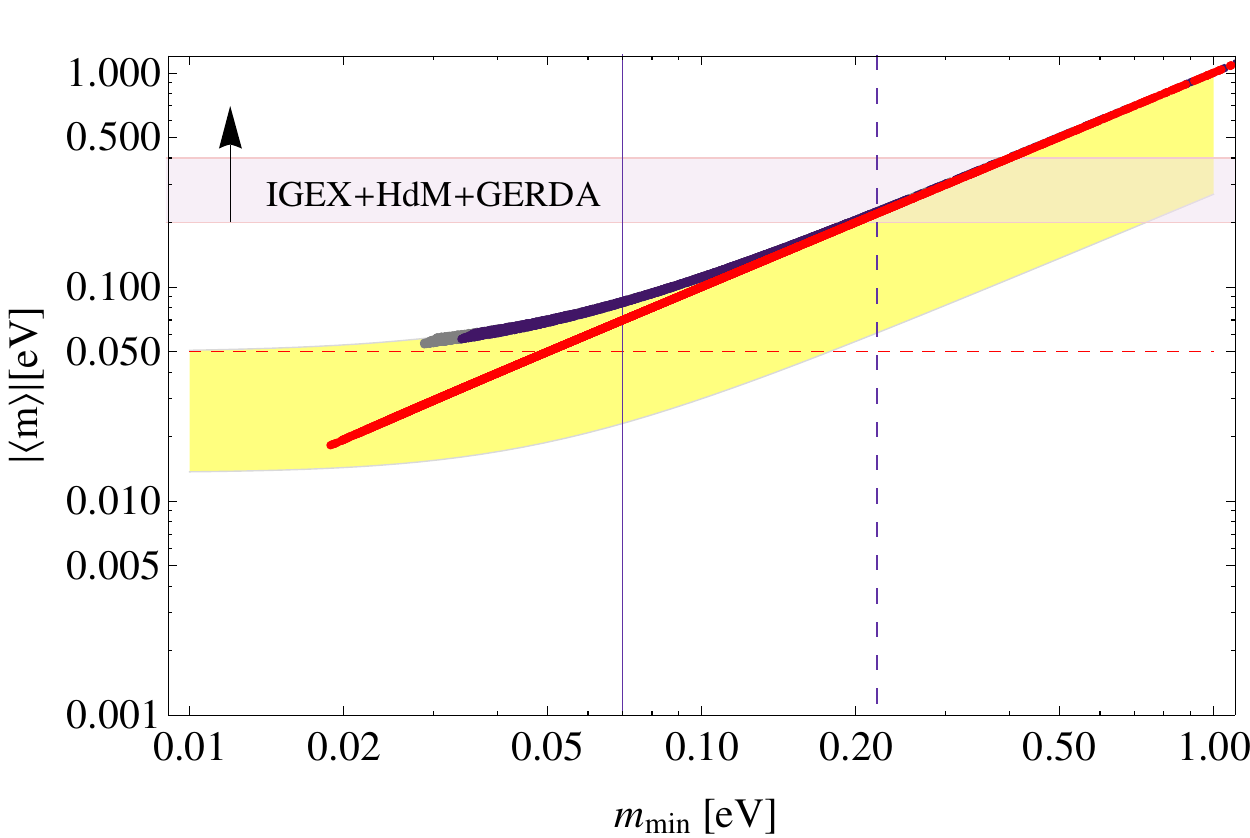}}
     \end{center}
\vspace{-1.0cm} \caption{\it The plots show \meff as a function of
the lightest neutrino mass for NO (Left Panel) and IO (Right Panel)
using 3$\sigma$ uncertainty in the oscillation data. The upper plots
refer to $B_1$ and $B_2$ textures, the lower ones to $B_3$ and
$B_4$. We used different colors to distinguish the textures, given
as follows: $B_1$ in orange, $B_2$ in violet, $B_3$ in gray, $B_4$
in black, $C$ in red (only for IO). The solid vertical line is the
Planck limit on $m_{min}=0.07$ eV, whereas the dashed one represents
the value $m_{min}=0.22$ eV implied by the more conservative
constraint $\Sigma < 0.66$ eV. The region above the horizontal lines
(as indicated by the arrow) is now excluded, according to the
analysis presented in \cite{Agostini:2013mzu}. The dashed horizontal
line represent future sensitivity for the \betabeta-decay process.
\label{fig:meff}}
\end{figure}

In Fig. \ref{fig:meff} we show for the textures under study the
values of \meff versus the lightest neutrino mass in the  NO (Left
Panel- B textures only) and IO (Right Panel- all textures)
considering 3$\sigma$ uncertainty in the oscillation parameters.
Since the regions identified for the textures of type B are
partially overlapping, we display $B_1$ and $B_2$ in the upper plots
and $B_3$ and $B_4$ in the lower ones. It is clear from the plots
that part of the solutions are strongly disfavoured (especially in
the IO case) by the upper limit $m_{min}\lesssim 0.07$ eV implied by
eq.(\ref{eq:PlanckBAO}) (solid vertical line); the recent combined
\betabeta-decay limit $T^{1/2}_{0\nu}(^{76}Ge)> 3.0\times 10^{25} $y
(90\% CL) \cite{Agostini:2013mzu}, corresponding  to minimal values
of \meff in the range $(0.2-0.4)$ eV, is less effective in this
respect.\footnote{The band of minimal values are obtained
considering as standard mechanism the exchange of a light Majorana
neutrino and different sets of nuclear matrix elements.} From  Fig.
\ref{fig:meff} one can also appreciate that even using a more
conservative limit given by the Planck Collaboration on the sum of
the neutrino masses, i.e. $\Sigma<0.66$ eV, implying
$m_{min}\lesssim 0.22$ eV (dashed vertical line in the plots) part
of the solutions found for the textures of type $B$ and $C$ in the
case of NO and IO are now excluded.

In the light of the Planck results, we find particularly interesting
to study with some details the correlations implied by the textures
$B_{1-4}$ and $C$ between the neutrino masses (and their sum) and
the atmospheric angle, the worst known mixing angle up to now. In
order to do that, it is necessary to express the neutrino masses in
terms of $\theta_{23}$; referring to the approximate results
reported in \cite{Fritzsch:2011qv} (still good with the mixing
parameter values of Tab.\ref{tab:neu}), we have:
\begin{itemize}
\item Textures $B_1$ and $B_3$ (for NO or IO):
\be m_1 \approx m_2 \approx m_3\tan^2\theta_{23}, \quad   m_3
\approx  \sqrt{ \frac{\Delta m^2_{A}}{1-\tan^4\theta_{23}}}, \ee

\item Textures $B_2$ and $B_4$ (for NO or IO):
\be m_1 \approx m_2 \approx m_3\cot^2\theta_{23}, \quad   m_3
\approx \sqrt{ \frac{\Delta m^2_{A}}{1-\cot^4\theta_{23}}}, \ee

\end{itemize}
where $\Delta m^2_{A}=\Delta m^2_{31(32)}$ for NO (IO).
Texture $C$ is a bit more complicated since the parameter space for NO is reduced to values
of $\theta_{23}$ very close to maximal mixing whereas the one for IO is much larger.
In the latter case, taking full advantage of the analytical
results presented in \cite{Fritzsch:2011qv},
we get:
\bea
m_3 &\sim&  \sqrt{\Delta m_A^2}\,s_{13}\,\cot \theta_{12}\, |\tan 2\theta_{23}|\, (1- s_{13}\,\cot \theta_{12}\,\tan 2\theta_{23})\;,\cr
m_2 &\sim& \sqrt{\Delta m_A^2}\,|\tan 2\theta_{23}|\,\left(\frac{s_{13}^2\,\cot^2 \theta_{12}+2\,\cot^2 2\theta_{23}}{2\,|\cot 2\theta_{23}|}\right)\;,\label{massesC}\\
m_1 &\sim& \frac{\sqrt{\Delta m_A^2}\,|\tan 2\theta_{23}|}{2\,|\cot
2\theta_{23}|}\, \left[2 \cot^2 \theta_{12} \cot^2 2\theta_{23} \nn
-2 s_{13} \cos \delta \cot \theta_{12}\cot 2\theta_{23}\csc^2
\theta_{12}+ \right.\\ &&  \nn \left.s_{13}^2 \left(1+\cos^2 \delta
(1+2\cos 2\theta_{12})\csc^4\theta_{12}\right)\right]\;. \eea 
All the previous expressions are obtained for small
$\theta_{13}$; however, we checked that they are valid within 10\%
with respect of the exact values obtained using the best fit values
of Table \ref{tab:neu}.

In the NO, on the other hand, the ratios among the complex neutrino masses
$\lambda_i$ computed for $\theta_{23}=\pi/4$ are given by: \bea
\frac{\lambda_1}{\lambda_3}=\frac{\lambda_2}{\lambda_3} =
-\frac{c_{13}^2}{1+s_{13}^2\,e^{2 i
\delta}}\,, \eea which gives degenerate $m_1$ and $m_2$. In
order to reproduce the solar squared mass difference, the
atmospheric angle must depart from its maximal value by very small
quantities, otherwise the ratio $r\equiv \Delta m^2_{21}/\Delta
m^2_{31}$ would be too large. This points towards a very small
dependence of the neutrino masses on $\theta_{13}$, so as for
$\Sigma$. 

We are now in the position to express $\sin^2 \theta_{23}$ as a function of $\Sigma$;
approximate expressions are the following:
\begin{itemize}
\item Textures $B_1$ and $B_3$
\be \sin^2\theta_{23}= \left(- \Sigma^2 -\Delta m^2_{A} +\Sigma
\sqrt{\Sigma^2+ 3\Delta m^2_{A}}\right)/\Delta m^2_{A}\label{B1}\ee
\item Textures $B_2$ and $B_4$
\be \sin^2\theta_{23}= \left( \Sigma^2 +2\Delta m^2_{A} +\Sigma
\sqrt{\Sigma^2+ 3\Delta m^2_{A}}\right)/\Delta m^2_{A}\;.\label{B2}\ee
\end{itemize}

\begin{figure}[h!]
  \begin{center}
 \subfigure
 {\includegraphics[width=7cm]{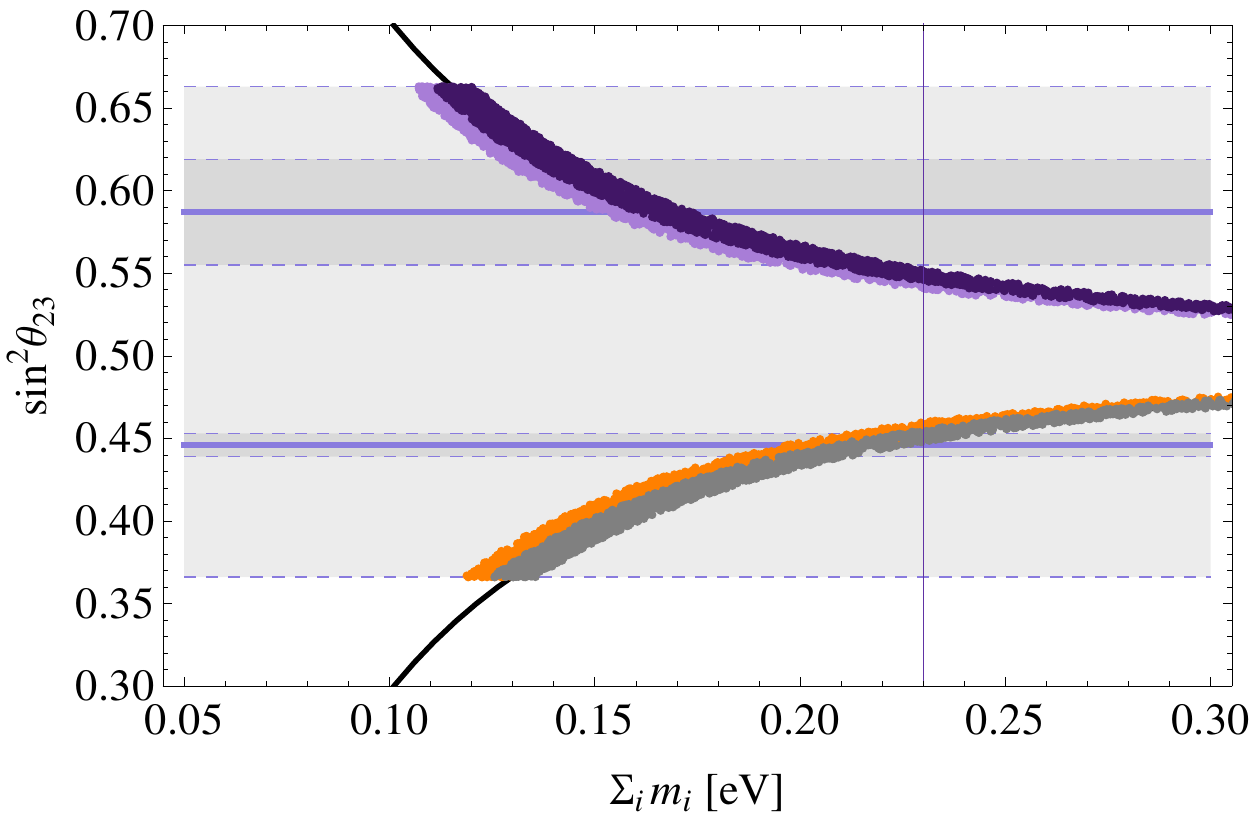}}
 \subfigure
   {\includegraphics[width=7cm]{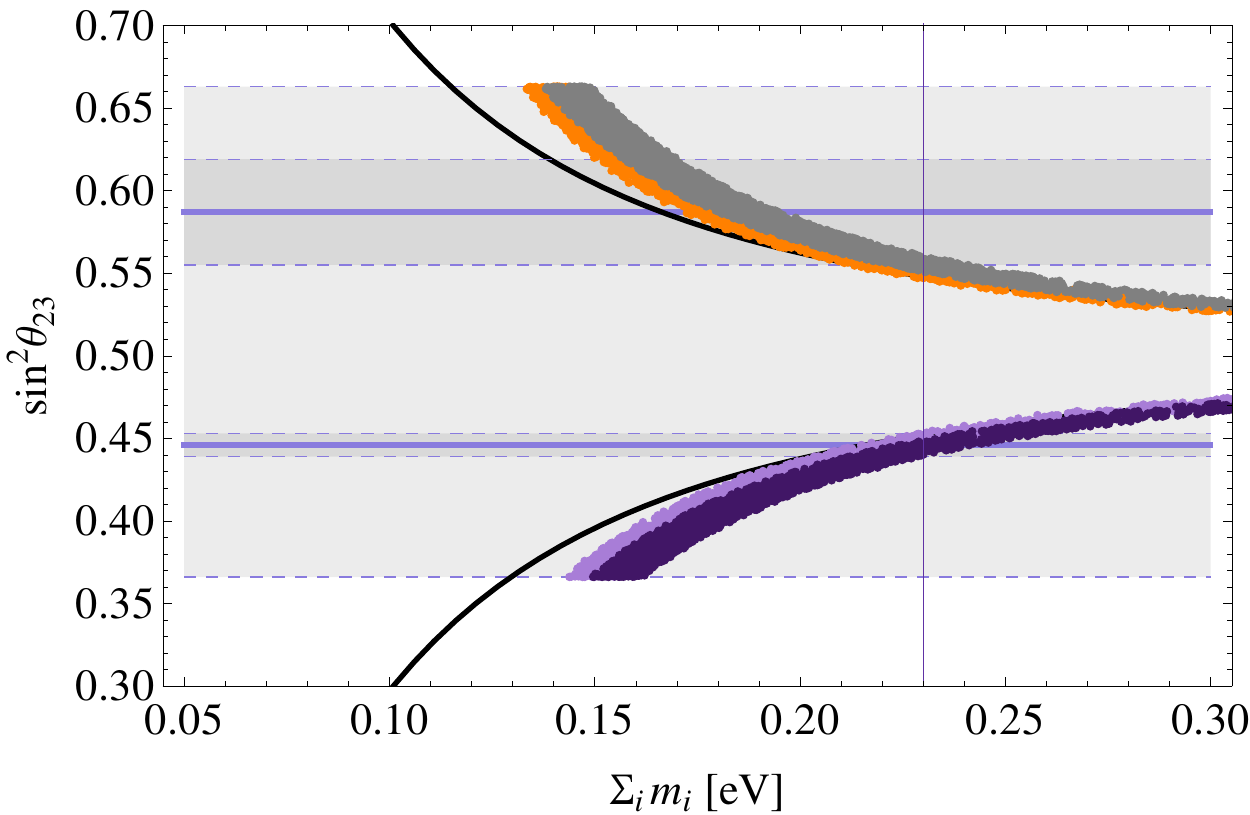}}
     \end{center}
\vspace{-1.0cm} \caption{\it The plots show the correlation between
$\sin^2\theta_{23}$ and the sum of the light neutrino masses for NO
(Left Panel) and IO (Right Panel).  The Textures are given as
follows: $B_1$ in orange, $B_2$ in violet, $B_3$ in gray, $B_4$ in
black, $C$ in red (only for IO). The area on the right  of the solid
vertical line represents the constrain of eq. (\ref{eq:PlanckBAO})
given by Planck \cite{Ade:2013lta} and is therefore strongly
disfavoured while the two strong shaded horizontal bands corresponds
to 1$\sigma$ uncertainty in the octant degeneracy in the
determination of $\theta_{23}$. Solid black lines represent the
approximate expressions of eqs.(\ref{B1}) and (\ref{B2}).
\label{fig:sumvst23} }
\end{figure}

In Fig. \ref{fig:sumvst23} we show the numerical correlation between
$\sin^2\theta_{23}$ and $\Sigma$ for the textures of type B analyzed
in this paper. The horizontal light and the strong shaded band
correspond respectively to the 3$\sigma$ and 1$\sigma$ uncertainties
in the mixing angles and two squared mass differences while the
solid vertical lines correspond to the Planck limit of eq.
(\ref{eq:PlanckBAO}). The solid black lines represent the
approximate expressions given above for the $\theta_{23}-\Sigma$
correlations, eqs. (\ref{B1}) and (\ref{B2}).

As already indicated by the analysis of the textures at the best fit
values, one can see in the left (right) panel that the textures
$B_1$ and $B_3$ ($B_2$ and $B_4$) are compatible only with the value
of $\theta_{23}$ in the lower (upper) octant. Given the 3$\sigma$
ranges on $\theta_{23}$, the Planck limit already rules out part of
the correlations implied by the four textures, for both NO and IO. A
closer look at the 1$\sigma$ level, however, shows more interesting
results; in fact, for the NO case (left panel), the Planck limit
does rule out part of the correlation predicted by the textures
$B_{1,3}$ (thus, only if $\theta_{23}$ will result to be smaller
than maximal mixing, as obtained in \cite{Capozzi:2013csa}). A
similar situation also happens in the IO case (right panel) for the
textures $B_{2,4}$. We stress again that at the 3$\sigma$ level no
definite conclusions can be drawn.

For texture $C$, in the NO case and for maximal $\theta_{23}$ we get
a simple expressions for $\Sigma$: \bea \Sigma = \sqrt{\Delta
m_A}\,\frac{\left[3- s_{13}^2(2-\cos 2\delta )\right]}{2 s_{13}|\cos
\delta|} \,. \eea This expression has a minimum value corresponding
to CP conserving values of $\delta = (0, \pi)$; for the intervals of
the mixing parameters reported in Tab.\ref{tab:neu}, this gives:
\bea \Sigma_{min} \sim 0.45\, \mbox{eV}. \eea This is the main
result of the paper, in that the Planck result on the sum of the
neutrino masses completely excludes the texture $C$ as a viable
Majorana mass matrix to describe the neutrino properties. For the IO
case the expression of the masses in eq. (\ref{massesC}) shows a
complicate dependence on $\theta_{23}$, so we prefer to give
$\Sigma$ as a function of $\theta_{23}$ and not viceversa; we get:
\bea \Sigma &=& \frac{\sqrt{\Delta m^2_{21}}}{2}\, \left[ 2 \csc^2
\theta_{12}-2 s_{13}\cot\theta_{12}\tan 2\theta_{23} (-1+\cos \delta
\csc^2 \theta_{12}) + \right.\nn \cr &&
 \left.  s_{13}^2\,(-2 \cos \delta \cot^2\theta_{12}+\csc^2 \theta_{12}+\cos^2\delta (1+2 \cos 2\theta_{12})\csc^4 \theta_{12}\tan^22\theta_{23})
\right].
\eea
\begin{figure}[h!]
  \begin{center}
 \subfigure
    {\includegraphics[width=7cm]{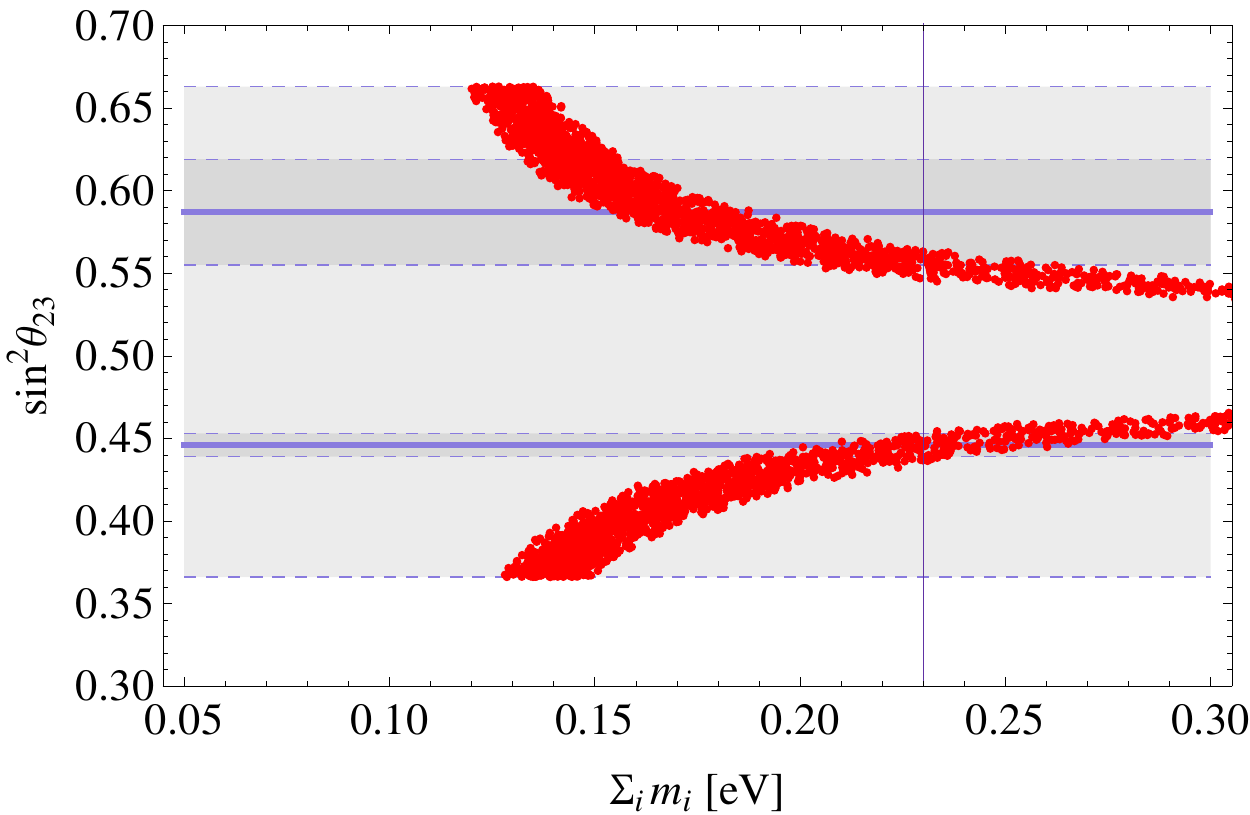}}
 \subfigure
 {\includegraphics[width=7cm]{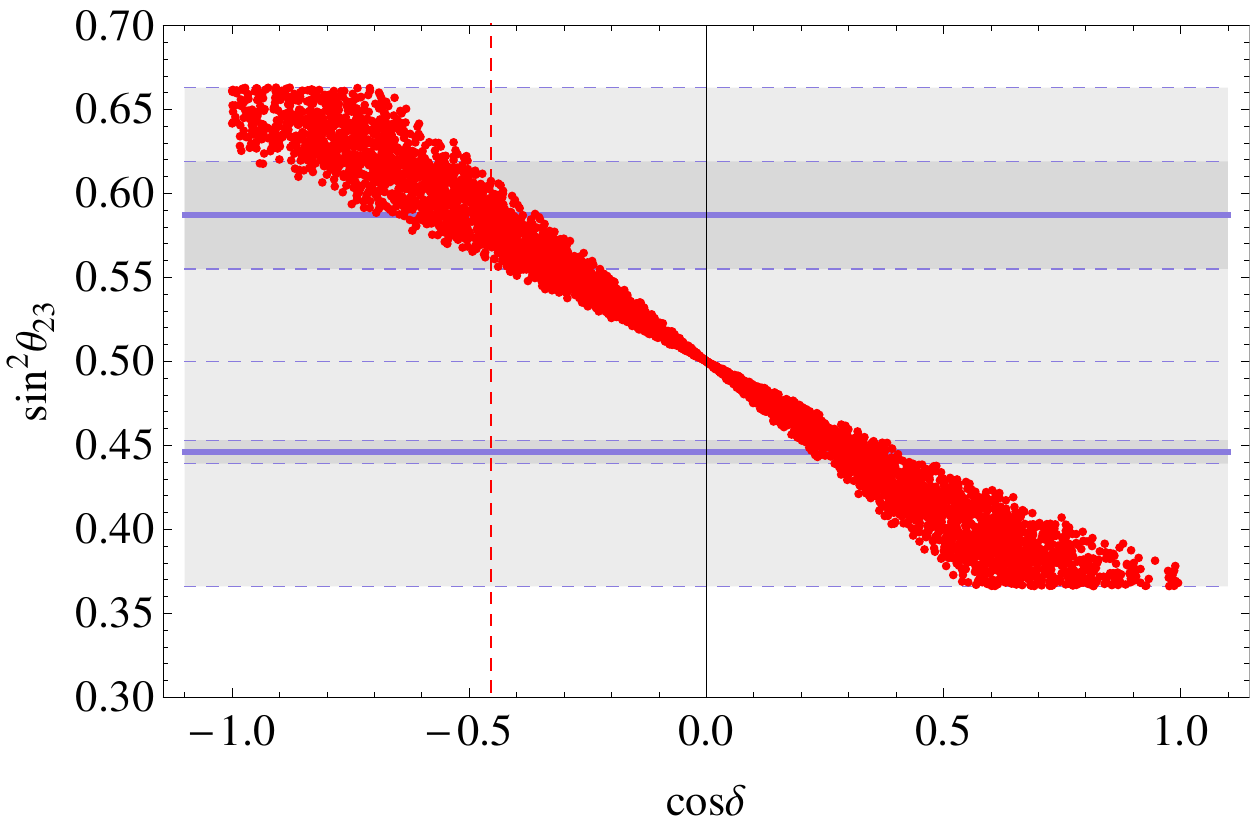}}
     \end{center}
\vspace{-1.0cm} \caption{\it (Left Panel) The plot shows the correlation in IO between
$\sin^2\theta_{23}$ and the sum of the light neutrino masses for the texture $C$.
The area on the right of the solid vertical line is strongly disfavoured by
the Planck data. As in Fig. \ref{fig:sumvst23}, the strong shaded area corresponds to
the 1$\sigma$ uncertainty in the determination of $\theta_{23}$.
(Right Panel) We show the corresponding correlation between
$\sin^2\theta_{23}$ and $\cos\delta$. The dashed vertical line corresponds to the
preferred best fit values for $\delta$ given in the global fit analysis of Ref. \cite{Capozzi:2013csa}.
\label{fig:sumvst23C} }
\end{figure}
In Fig. \ref{fig:sumvst23C} we show in the left panel the
correlation between the $\sin^2\theta_{23}$ and $\Sigma$ for the
texture $C$ in the case of IO. We notice that the values of
$\theta_{23}$  selected in our numerical analysis for IO do not
prefer any specific octant and the same considerations of the type B
textures apply. However, if one considers the 1$\sigma$ uncertainty
in the octant degeneracy, the numerical solutions we found for the
lower value of $\theta_{23}$ are constrained in a narrow interval
corresponding to values for the $\Sigma$ very near to the Planck
limit while a 1$\sigma$ uncertainty in the upper value of
$\theta_{23}$ is compatible with a broader area. It is interesting
to observe that the correlation between the value of $\delta$ and
$\sin^2\theta_{23}$ is such that the best fit for $\delta$ of
\cite{Capozzi:2013csa} (dashed vertical line in the  right panel of
Fig. \ref{fig:sumvst23C}) points towards a second octant solution
for $\theta_{23}$, already at the 1$\sigma$ level.

In conclusion, we have reanalized some of the two-zero neutrino
textures in the light of the recent Planck result on the masses of
the light neutrinos and we have shown that texture $C$ is not
compatible with the normal ordering of the neutrino mass
eigenstates. As a byproduct, we have also shown that, taking the
atmospheric angle with a 1$\sigma$ uncertainty and in the first
octant, strong constraints are put on the matrices of $B$ types,
which allow to disfavor $B_{1,3}$ in NO and $B_{2,4}$ in IO.

\subsection*{Acknowledgments}
D.M. and A.M. acknowledge MIUR (Italy) for financial support under
the program ``Futuro in Ricerca 2010'', (RBFR10O36O).

\end{document}